# Beyond-limit light focusing in the intermediate zone

K. R. Chen[1,2,*], W. H. Chu[3], H. C. Fang[3], C. P. Liu[3], C. H. Huang[2], H. C. Chui[2], C. H. Chuang[4], Y. L. Lo[4], C. Y. Lin[1], S. J. Chang[5], F. Y. Hung[6], H. H. Hwuang[7] and Andy Y.-G. Fuh[1,2]

*[1]Department of Physics, [2]Institute of Electro-optical Science and Engineering, [3]Department of Materials Science and Engineering, [4]Department of Mechanical Engineering, [5]Department of Electrical Engineering, [6]Institute of Nanotechnology and Microsystems Engineering, [7]Department of Hydraulic and Ocean Engineering, National Cheng Kung University, 1 University Road, Tainan 70101, Taiwan, ROC.*

**Abstract**

We experimentally demonstrate that a new nanolens of designed plasmonic subwavelength aperture can focus light to a single-line with its width beyond the diffraction limit that sets the smallest achievable line width at half the wavelength. The measurements indicate that the effect of the near-field on the light focused is negligible in the intermediate zone of $2 < kr < 4$ where the line-width is smaller than the limit. Thus, as a verification of theoretical prediction, the fields focused are radiative and with a momentum capable of propagating to the far zone as concerned by the limit.

Nano-photonic metamaterials have unusual properties such as negative refraction [1-2] and extraordinary light transmission [3-4]. A superlens [5-10] made of negative index materials (NIM) was proposed [5] to amplify the evanescent near-field [11-14] in order to image an electrostatic field source with its size smaller than half the illumination wavelength, $\lambda$. In an implementation [6], as a typical example, the object layer of the sandwiched superlens is with metal lines of $\lambda/6$ width. The electrostatic field of the line charges polarized by the illuminating light excites the surface plasmons on the Ag film to produce an image on the surface of the adjacent photoresist layer. The half-pitch resolution of the photoresist image can remarkably be as small as the metal line width of $\lambda/6$ that is better than that of the conventional lens limited by E. Abbe's diffraction [15-16], but the typical averaged height modulation is less than $\lambda/30$. Furthermore, the physical mechanisms and the governing equations are different. For the superlens, what is involved is the electrostatic field governed by the Poisson equation. Also, both the object and the image have to be in the near zone [17]. Although the image's electrostatic near-field in another implementation [7] can be coupled to propagating fields outside the near zone, the full-width at the half-maximum (FWHM) becomes larger than that limited by Abbe's diffraction.

In general, there are two different definitions of the diffraction limit. Abbe's theory [15-16] of image formation and the equivalent approach by L. Rayleigh [15,18] concern the resolution of two light spots resulting from the focusing of two light sources by a lens. The propagating fields governed by Maxwell's wave equations are of concern by the diffraction limit. Thus, the superlens bypasses Abbe's imaging limit, but does not challenge the physical mechanism of the diffraction limit.

A more rigorous and influential definition is on the spot size of a light focused by a lens, defined as the FWHM of the wave energy or as twice the position uncertainty; the smallest line width achievable is half the wavelength. This limit is the ultimate manipulability and resolution of numerous diagnostic and fabrication instruments. This limit also inspired W. Heisenberg's quantum uncertainty principle [19] that is a foundation of modern science; in fact, they can be deduced from each other [19-21].

In contrast to the superlens, superoscillation [22-25] suggests that a wave function can vary spatially faster than the highest frequency component. Interestingly, a quasi-crystal array of about 14 000 nanoholes [23-24] of 200 nm diameter in a metal screen produces a pattern of hot spots. However, the distance between the spots is larger than Abbe's limit, and the size of the overall wave function remains as large as the array area of 0.2 mm diameter. Also, generating a single focal spot involves difficulties.

In this letter, an innovative approach and the mechanisms to focus light to a single-line with its width beyond the diffraction limit [26-27] are experimentally demonstrated with a nanolens including a metallic film with a double-slit and a patterned exit structure, as shown in Fig. 1. The focused light is measured by a near-field scanning optical microscope (NSOM) and an optical microscope (OM). The results verify that the focused fields are radiative and the involvement of near-field is negligible [26-27].

A thin silver film on the fused silica substrate fabricated by a Focused Ion Beam (FIB) (as shown in Fig. 1b and 1c) is employed as our lens of focusing aperture beyond (FAB) the diffraction limit. The refractive index of the silver film is $0.13401 + i\, 3.9704$ for 630.6 nm and $0.13614 + i\, 4.0342$ for 638.3 nm from ellipsometry (M44, J. A. Woolam) measurement. The substrate has a refractive index of 1.45840 for 589.3 nm and a dielectric constant of 3.80 at 20 °C, 1 MHz, according to the supplier (Corning).

The FIB milling starts with a trench of 480 nm in width and 80 nm in depth, followed by the slits and then the grooves.

In our experiments, a plane wave of 633 nm light illuminates the FAB nanolens. With the NSOM scanner (Aurora-3, Veeco) [28-29], the focused light is collected by tapered Al-coated fiber probes (1642-00, Veeco) 50 nm or 100 nm in diameter and then detected by a photomultiplier tube (R2949, Hamamatsu) through a microscope objective (PlanAPO, 50X, NA＝0.8, Olympus). The constant scanning rate in $x$-$z$ is set at 5 μm/s, and the scanning range is 1000×1000 nm$^2$. When using the OM (BX51M, MS Plan 100X objective, NA=0.95, Olympus), the focused light propagates about a distance of 500 λ to reach the object lens and then about 295,000 λ further for being recorded by the charge-coupled device. In the Finite-Difference-Time-Domain (FDTD) simulation [30], the system has 1000 × 500 cells of the Yee space lattice with a cell size of 5 nm. The refractive index used is 1.4568 for fused silica [31] and 0.13417 + $i$ 3.9915 for silver [31]. The bottom of the film is at the $y$ = 200 cell. The time is normalized to the light period, and the time step is 0.005.

The light with polarized electric $E_x$ and magnetic $H_z$ fields propagates upward through the double-slit. At the exit of each slit, the structure is not symmetrical. The fields are bent toward the center. Besides being transmitted, the $H_z$ field of subwavelength scale can be produced at the central area (where the interference of the $H_z$ field is constructive) by the current on the metal strip surface and the conversion effect. The $E_y$ field, the polarized surface charge and the time-averaged Poynting vector in the $x$ direction of the bent and diffracted light from one slit are cancelled with those from the other. The cancellation converts their energies to the focused $H_z$ field.

Figure 2a shows the local field intensity of the focused light measured by the NSOM with a 50 nm probe. The homogeneity of the measurement profile along the $z$ direction is good enough to evidence the quality of the film and the structure. The FWHM of the profile across the slit structure in the $x$ direction is obviously smaller than half the wavelength and thus, the diffraction limit. The experimental profile averaged over the $z$ direction and that calculated from the FDTD simulation are shown on Fig. 2b; they are found to be in a good agreement. The FWHM of the FDTD simulation is 0.287 $\lambda$, while the NSOM measurement gives 0.34 $\lambda$. The position uncertainty, defined as $\Delta x = (<x^2>-<x>^2)^{1/2}$, where $<f> = \int H_z^2 f\, dx / \int H_z^2\, dx$, calculated by the simulation is 0.109 $\lambda$, averaged over the focused line; it is 0.177 $\lambda$ when over the profile. The width of the $H_z$ energy averaged-along-$x$ is twice the position uncertainty and thus is 0.218 $\lambda$ over the focused line (or 0.354 $\lambda$ over the profile). All the widths shown are smaller than the width of the central metal strip, half the wavelength, and thus are beyond the limit. Obviously, the diffraction limit has been surpassed by the result of this FAB lens, in which there is such a small single-line width occurring with regard to the focused light.

Figure 3a indicates that the focused light can propagate out and can be measured by the OM located far away from the FAB lens. Figure 3b shows that the time-averaged Poynting vector in the original $y$ propagation direction is not zero in contrast to that of near-field. Poynting vector represents the momentum of propagating light fields. Thus, this confirms that the focused fields have the momentum to propagate to the far zone [17]. The ability of the focus fields to propagate is of academic concern with regard to the diffraction limit. Since this data is an indicator of the capability and simplicity of this approach for moving the focal point and the field energy away from the surface, it is clear that the FAB nanolens is superior to evanescent near-field techniques [5-14] for many critical applications.

The involvement of near-field on the line width of the light focused is quantitatively investigated in-depth. Figure 4a indicates the agreement of the NSOM measurements and the simulation results on the FWHM is good for the normalized distance $2 < kr < 4$ (i.e., between 1/3 and 2/3 of the wavelength.) At the low $kr$ region, the time-averaged $E_x$ field energy peaks at the two converging paths of the bended $E_x$ field from the two slits so that it is only plotted for the larger $kr$ region. The FWHMs of both the calculated snapshot $E_x$ and $H_z$ field energies are found to agree well with the FWHMs of the time-averaged $E_x$ and $H_z$ field energies. As for the NSOM measurements, the smaller the width of the probes used, the smaller the NSOM measured line width of the focused light. The increase of the FWHMs of the NSOM measurements at a low $kr$ may be caused by the near-field close to the metal strip, the mechanism of the propagating field collection by the probe and their coupling; in other words, the FWHM of the near-field in the near zone may be larger than that of the focused light in the intermediate zone so that it is unlikely to reduce the line width. The near-field decreases when the probes move away from the surface. From the NSOM measurement outside the focused area at $x = 320$ nm (Fig. 4b), which is dominated by the near-field, the distance of half the measured field energy is $kr = 0.66$ or $0.1\ \lambda$; that is consistent with the near zone boundary measured by other NSOMs [28-29] and also with our measurements (Fig. 4b) at $x = 0$ nm (i.e., at the center of the aperture). The profiles of our NSOM measurement at the center, including the peak locations, are found to agree with those of the time-averaged $E_x$ field energy from the simulation. The derivative of the NSOM measured profiles changes its sign at $kr = 1$ so that the increase of the propagating field amplitude is larger than the decrease of the near-field amplitude at $kr > 1$; in other words, the amplitude of the near-field is smaller than that of the propagating field at $kr > 1$. The NSOM measurement results indicate that the effect of the near-field is negligible in the intermediate zone of $2 < kr < 4$, as also verified by the simulation. In this intermediate zone, the line width of the light focused by this FAB nanolens is below that of the diffraction limit.

The FAB nanolens capable in focusing light beyond the diffraction limit in the intermediate zone is intellectually intriguing and important for application possibilities. For example, as limited by the diffraction, nanofabrication with photolithography is a key issue preventing the further progress of the semiconductor industry according to Moore's Law. Any practical solution for nanofabrication with photolithography is required to have a finite depth of focus; the thickness of the exposed photoresist needs to be larger than the half-pitch resolution in order to fabricate the dots and wires of a circuit. A finite working distance between the lens and the photoresist is required to accommodate the surface topology of the mask and the photoresist, as well as the gap in a proximity mode. Obviously, these requirements are more easily satisfied by a FAB nanolens than by a superlens. As for bio-photonic applications, this FAB nanolens and approach enable unique opportunities. With a low-intensity light, individual subwavelength non-destructive nano-imaging and manipulation become possible inside a living cell or biological specimen. So, there is no need to rely on specific molecular absorption resonances. Besides imaging and manipulation, possible applications include probing the structure and dynamics of biological and physical systems at a smaller scale, the diagnosis and modification of material surfaces with greater precision, the squeezing of light into photonic and plasmonic circuits [32], and the connection of optical systems and finer electronic circuits, among many others.

In summary, a miniature FAB nanolens of subwavelength structure experimentally demonstrates the focusing of light in the intermediate zone to a single-line with its width beyond the limit of diffraction. Besides the academic interest generated by this nanolens and approach on surpassing the fundamental physics limitation, nano-photonic applications are self evidenced, especially with regard to the capability of reducing the sizes of the focused light spot and the device and of moving the sub-limit light spot outside the near zone and away from the surface.

**Acknowledgements** This work was supported by NCKU Project of Promoting Academic Excellence & Developing World Class Research Centers.

*chenkr@mail.ncku.edu.tw

## Figure Captions

Fig. 1 (Color online) The aperture structure and approach. (a) Schematic diagram of the aperture structure and the paths of the light transmitted, bent, generated and focused. (b) and (c) The scanning electron microscopy images revealing the top (b) and 52 degree tilt cross-sectional (c) views of the aperture (slit length, 4 μm) using a silver film on fused silica substrate, fabricated by the FIB; the Pt shield is not part of the FAB nanolens and was added to protect the lens solely for the purpose of making this cross-sectional view. The red dot lines define the silver film structure (slit width, 80 nm; groove width and depth, 80 nm; slit-groove distance, 160 nm; film thickness, 200 nm; central film thickness, 120 nm, width, 320 nm) also used for the simulation with the coordinate shown.

Fig. 2 (Color online) The field profiles from the NSOM measurements and the simulation. (a) the *x-z* distribution of the NSOM measurement. (b) The profiles of the NSOM measurement (green) and the peak focused $H_z$ field energy (red) from the simulation. The NSOM profile is from the *x-z* distribution averaged over *z*, and its peak is normalized to that of the simulation.

Fig. 3 (Color online) The experimental and simulation results which verify the propagation of the focused light. (a) The picture taken by the OM measurement; the image size shown is 100 × 240 pixels taken using a digital camera with the effective image resolutions in 4080 × 3072 (12.5 megapixels). (b) The time-averaged contours of the Poynting vector in the *y* direction.

Fig. 4 (Color online) The $r$ profiles of the fields from the NSOM measurements and the simulation, where $r$ is the $y$ distance from the metal surface. (a) The FWHM vs. normalized $r$ profiles of the NSOM measurements with the probe diameter of 50 nm (dark green dots) and 100 nm (light green triangles), the snapshot $H_z$ field energy (red dots), the time-averaged $H_z$ field energy (red curve), the snapshot $E_x$ field energy (blue dots) and the time-averaged $E_x$ field energy (blue curve). The error bars of the NSOM measurements are smaller than the size of the dots. (b) The $r$ profiles of the NSOM measurements at $x = 320$ nm (purple squares), at $x = 0$ nm with the probe diameter of 50 nm (dark green dots) and 100 nm (light green triangles), and the time-averaged $E_x$ field energy (blue curve) normalized to the peak of 50 nm NSOM measurement.

Fig. 1

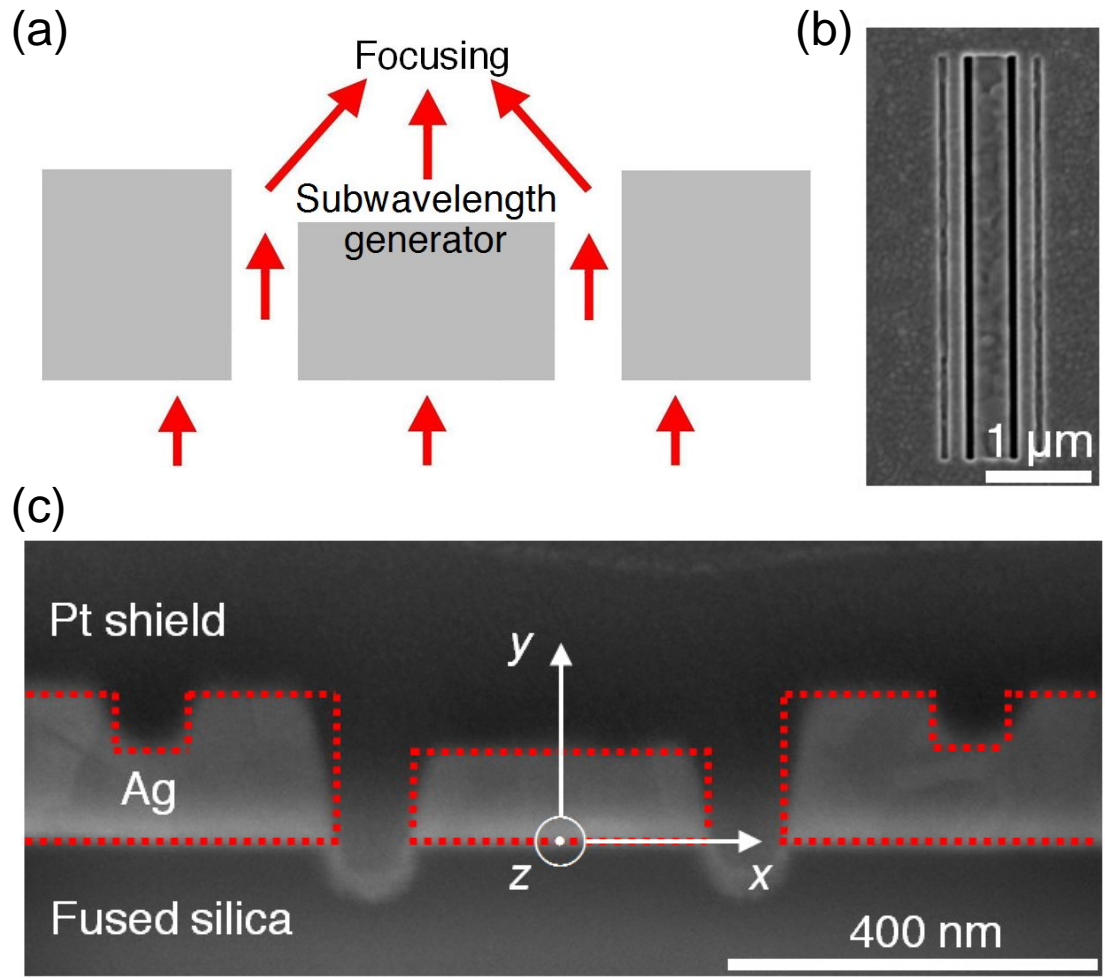

Fig. 2

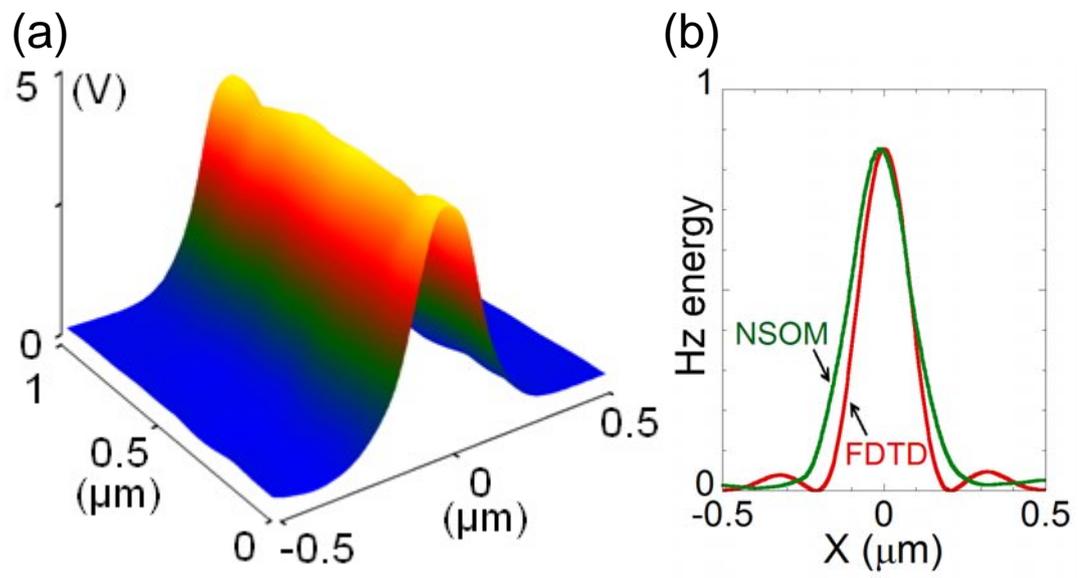

Fig. 3

(a) 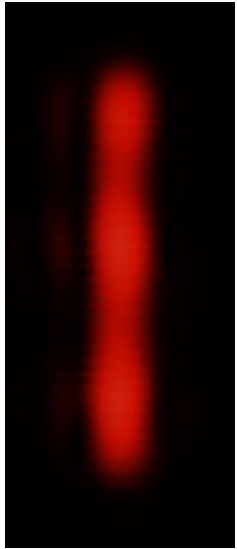 (b) 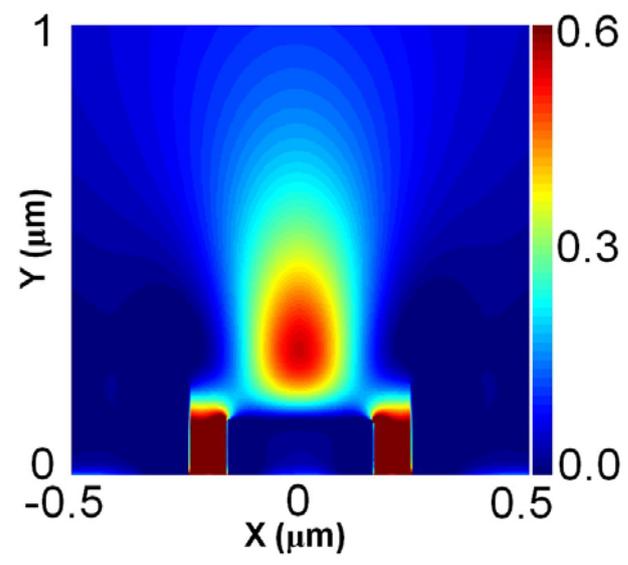

Fig. 4

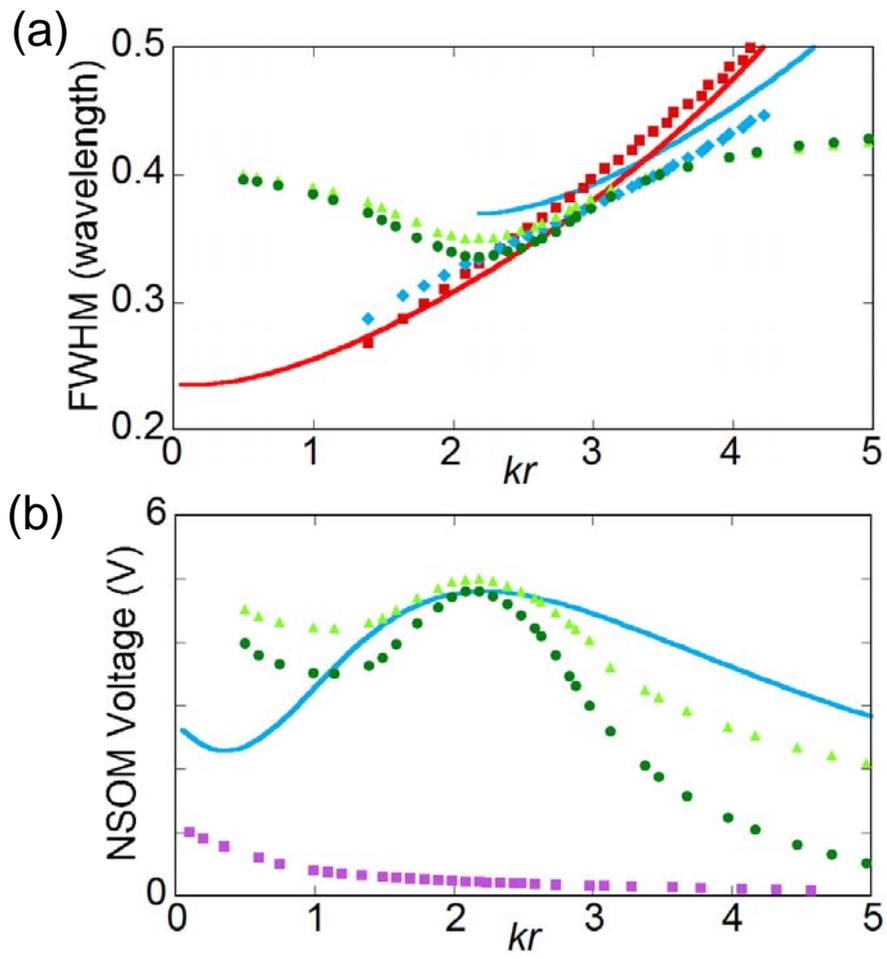